\documentclass[aps,nofootinbib,floats,superscriptaddress]{revtex4-2}

\usepackage{graphicx}
\usepackage{epstopdf}
\usepackage{amssymb}
\usepackage{amsmath,bm}
\usepackage{psfrag}
\usepackage{epsfig}
\usepackage{float}
\usepackage{bm}
\usepackage{textcomp}
\usepackage{hyperref}

\begin{document}
\title{
Minimal model of a 
Plateau-Rayleigh instability of a solid binary alloy nanowire: impacts of surface segregation 
}

\author{Mikhail Khenner\footnote{Corresponding
author. E-mail: mikhail.khenner@wku.edu.}}
\affiliation{Department of Mathematics, Western Kentucky University, Bowling Green, KY 42101, USA}
\affiliation{Applied Physics Institute, Western Kentucky University, Bowling Green, KY 42101, USA}

\begin{abstract}
\noindent
    
Minimal model and its linear stability analysis (LSA) for Plateau-Rayleigh (PR) instability of an 
intermixed, solid binary alloy nanowire is presented. Model accounts only for the surface diffusions 
of the alloy components without anisotropies of a surface energies, and for a classical thermodynamic surface segregation. 
Model allows to assess a deviation of a longwave instability cutoff wavelength $\lambda_c$ from the 
classical value $2\pi r_0$ (a wire circumference) as a function of a surface composition deviation 
from its equilibrium value, a wire radius, and a bulk concentration of a minority component. 
Results of LSA are presented for AuNi, PtNi, and CoNi nanowires, where an important distinction is that a 
surface segregation curve for CoNi is qualitatively different from the one for AuNi and PtNi. For each  
nanowire there is a sizable region 
of a parameter space where $\lambda_c<2\pi r_0$ and 
correspondingly, a fastest growing perturbation wavelength $\lambda_{max}$ is also less than a wire circumference. 
This is in stark contrast to PR instability of a single-component nanowire transporting mass by surface diffusion - 
in this case $\lambda_c$ and $\lambda_{max}$ are larger than a wire circumference.
Strong surface segregation is primarily responsible for unusually small $\lambda_c$ and $\lambda_{max}$. Results suggest that 
a large number of small and closely spaced nanoparticles may emerge 
upon disintegration via annealing of a binary alloy nanowire.

\medskip
\noindent

\end{abstract}

\date{\today}
\maketitle


\section{Introduction}
\label{Intro}

Solid metal or semiconductor nanowires are vulnerable to a morphological instability that may result in disintegration of a nanonowire 
into a string of nanoparticles. The instability, driven by a high-temperature surface diffusion, is of a 
Plateau-Rayleigh (PR) type \cite{XLL,LBPDVC,KTBECKN,KTEBCKN,NDPRC,BBSN,EWCWGBCCT}. 
Theoretical and computational studies of this instability continue since the publication of Nichols \& Mullins 1965 paper 
\cite{NM2,C,BBW,CFM,McCVoorhees,WMVD,KDMV,GM,GMM,DKW,WTLZNSMH,GP,Gill,BB,Hussein,BGS,MyWire1,MyWire2,MyWire3}. Fairly recently, due to proliferation of a physical, 
on-substrate synthesis methods for semiconductor nanowires 
the focus has shifted to understanding the surface energy anisotropy-driven faceting of a 
core \cite{EWCWGBCCT,DW,WJBS,WK} and core-shell nanowires \cite{S,VSZ,VSZ1,ZVD}. 

This notice is motivated by a surprising fact: despite studies of the PR instability for wires that span decades, the 
influence of alloying on a PR instability of a solid micro- or nanowire remains obscure. Although experiments that we 
are not aware of may have been published, we are certain that there is no theoretical, modeling, or computational works in the 
available literature. 

Therefore, the goal of this paper is the development of a minimal model for a post-growth PR instability of a metal binary 
alloy nanowire and the linear stability analysis (LSA) of that model.  By a metal binary alloy nanowire we understand a nanowire 
that has a bulk made of a homogeneous, or close to a homogeneous, mixture of two metals after it was grown. 

Our interest in a metal binary alloy nanowires is due to modern applications in heterogeneous surface catalysis 
\cite{Song,ZPT}. 
Such nanowires are routinely chemically synthesized in large quantities 
in a liquid solution \cite{STMTFTMAL,LJW}.  Wet chemical synthesis studies that we are aware of do not hint at a post-growth 
instabilities. However, device applications require a nanowire transfer to a substrate, where it will be subjected 
to strong electric current and mechanical stress. These factors may trigger  a 
morphological instability or enhance a PR instability. 
Core-shell metal binary alloy nanoparticles may be fabricated on a substrate by solid state dewetting of a thin metal bilayers 
\cite{HWKS,ELARC,CTZB,CTZB1,AKR1,AR}. 
Nanoparticles may result from disintegration of nanowires, but good intermixing in the bulk is difficult to accomplish, since 
the technique rests on a sequential deposition of planar layers of pure metals and therefore the diffuse interface between 
the layers is present during dewetting.
Also, recently a few advanced substrate-based vapor-liquid-solid like synthesis methods were developed for metallic and intermetallic 
binary alloy nanowires \cite{WLYHSZ,PMCCPB}.
Therefore, we consider a situation where a metallic binary alloy nanowire is on a substrate, where it is heated (annealed) 
allowing a PR instability to emerge.      

We base our minimal model on the continuum large model for solid binary alloys pioneered by Spencer \emph{et al.} \cite{SVT} and further developed by 
Zhang \emph{et al.} \cite{ZVD}, with contributions by Shenoy \cite{S} and Vastola \emph{et al.} \cite{VSZ,VSZ1}. That model, to our knowledge the only one 
of sufficient generality, was developed for growth of semiconductor 
binary alloy thin films, nanowires and nanoparticles as the primary application area. LSA is omitted in Refs. \cite{ZVD,S,VSZ,VSZ1}, where focus is 
on computations, whereas in Ref. \cite{SVT} LSA is presented for the case of a strained planar film growth. This setup is very 
different from the situation of interest in this paper, thus comparisons or analogies cannot be drawn. 
Moreover, despite that surface segregation is the major composition effect in thin alloy films \cite{WyK,Seah,RSK,LZH,AKR1,AR}, its impacts
on morphological instabilities are not well understood. Shenoy's and Vastola's \textit{et al.} models \cite{S,VSZ,VSZ1} 
include growth, faceting and bulk diffusion in addition to surface segregation. On the levels of either physics or model equations, this does not 
allow to fully grasp the mechanism and impact of surface segregation on morphological instability.  Therefore, our minimal model accounts only 
for the surface diffusions of two atomic species and a thermodynamic surface segregation. 


Lord Rayleigh's analysis, based on minimization of
a surface energy, confirmed Plateau's observations of a liquid jet and predicted, by identifying the fastest growing mode, that a liquid jet would
break up into droplets with a separation of $\lambda_{max} = 9.016 r_0\approx 2\pi \sqrt{2} r_0$, with $r_0$ the initial
radius of the jet \cite{Rayleigh}. The critical smallest perturbation wavelength for which 
a jet first becomes unstable is $\lambda_c=2\pi r_0$. This phenomenon is now known as PR instability. 
For free-standing solid rods (wires) evolving by surface diffusion with an isotropic surface energy, Nichols and Mullins \cite{NM2} 
showed that Rayleigh's instability criterion holds, i.e. all perturbations that have wavelengths larger than $\lambda_c=2\pi r_0$ 
destabilize a wire, and the fastest growing, i.e. the most dangerous wavelength is $2\pi \sqrt{2} r_0$. The factor of $\sqrt{2}$ 
is because the dispersion relation $\omega(k)$ of the instability is of a standard long-wavelength type (here, $\omega$ is a growth rate and $k$ a wavenumber). 
A spacing of approximately $\lambda_{max}$ between the particles, where $\lambda_{max}$ value $2\pi \sqrt{2} r_0$ is corrected due to 
presence of a substrate \cite{McCVoorhees}, polycrystallinity \cite{KTEBCKN}, or anisotropy \cite{EWCWGBCCT}  is indeed observed in experiments \cite{XLL,LBPDVC,KTBECKN,KTEBCKN,NDPRC,BBSN,EWCWGBCCT}.
Therefore a basic question for a binary alloy nanowire is whether alloying modifies the PR critical wavelength $\lambda_c$. 
In this notice we answer this question positively.

Interestingly, to our knowledge only two studies were published that address a PR type instability 
of a liquid binary mixture jet \cite{F} (a theoretical study) or of a liquefied, patterned binary alloy wires \cite{AKCRC} 
(a molecular dynamics simulation study). As the  titles of these papers suggest, the focus there is on understanding 
how phase separation and shear flow  \cite{F}, or phase separation  \cite{AKCRC} affect the PR wavelength. 
In contrast, our model is centered around the effect of a surface segregation and to a lesser degree, the kinetic effects 
due to the differences in the surface mobilities of the two components of the binary alloy.  
Of course, a liquid and a solid systems are only nominally similar in the first place, since even the major mass transport mechanisms are 
very different, i.e. a volumetric fluid flow and a surface diffusion, respectively.
 
For illustration of LSA results we chose three well-studied, Ni-rich binary alloy nanowires that are important for surface catalysis studies \cite{STMTFTMAL,LJW}, AuNi, PtNi, and CoNi at 650$^\circ$. 
At this temperature, Ni concentration higher than 70\%, and thickness/diameter of 100 nm or less, AuNi, PtNi, and CoNi thin 
alloy films and nanowires do not phase separate (the miscibility gap opens below roughly 550$^\circ$ \cite{LB,BKMK}) 
and the surface diffusion is five to seven orders of magnitude faster than bulk diffusion \cite{BR}.	
Values of needed physical parameters at 650$^\circ$ were obtained 
from standard databases \cite{SR,VRSK,NIST,DiffDatabase,NIST1}.

\section{The Model}
\label{Modell}

Let A designate atoms of a solvent (Ni), and B designate atoms of a solute (Au, Pt, or Co). 
It is assumed that the surface layer is uniform in composition, which is different from that of the bulk.
Our model for axisymmetric, surface diffusion-driven evolution of a wire surface $r=F(t,y)$, coupled to 
evolution of a surface concentration $C_B(t,y)$ of a solute, is based on a PDE system
\begin{eqnarray}
F_t&=&-\sqrt{1+F_y^2}\;{\mathbf \nabla}_s\cdot\left(\Omega_A{\bf J}_A+\Omega_B{\bf J}_B\right), \label{reqq}\\
\delta\frac{\partial C_B}{\partial t}&=&C_B^{(b)}{\mathbf \nabla}_s\cdot\left(\Omega_A{\bf J}_A+\Omega_B{\bf J}_B\right)-\Omega_B{\mathbf \nabla}_s\cdot{\bf J}_B +K\left(C_B^{(eq)}-C_B\right), \label{C_B_eqq}
\end{eqnarray}
where $y$ is the axial variable,  $-{\mathbf \nabla}_s\cdot\left(\Omega_A{\bf J}_A+\Omega_B{\bf J}_B\right)$ the 
normal velocity of a surface, ${\mathbf \nabla}_s$ the surface gradient operator in a cylindrical coordinates, $\Omega_A$ and $\Omega_B$ the atomic volumes of 
the two diffusing species in a binary alloy, ${\bf J}_A$ and ${\bf J}_B$ the surface diffusion fluxes of these species,
$\delta$ the surface layer thickness (typically 1 ML), $C_B^{(b)}$ a sub-surface concentration (i.e., in the bulk immediately below the surface layer), 
$C_B^{(eq)}$ the equilibrium surface concentration, and 
$K$ the exchange rate of material between the surface and the bulk. It measures
the strength of the system to equilibrate the surface composition toward its thermodynamic equilibrium state.
Eq. (\ref{reqq}) is a conventional 
PDE \cite{CFM,BBW,McCVoorhees,WMVD,SVT,ZVD} that describes a morphology evolution by surface diffusion of adatoms A and B. 
Concentration evolution equation (\ref{C_B_eqq}) is discussed in Appendix \ref{BasicEqs}.

Surface diffusion flux of $A$ or $B$ species is given by a surface gradient of a chemical potential times a concentration-dependent 
mobility \cite{SVT,ZVD}:
\begin{equation}
{\mathbf J}_i=\frac{-\Omega_i\nu_iD_i}{k T}C_i{\mathbf \nabla}_s \mu_i\equiv-M_iC_i{\mathbf \nabla}_s \mu_i,\quad i=A\ ,B. \label{J_i}
\end{equation}
We choose from now on to refer to a constant $M_i$ as the mobility, rather than to a product $M_i C_i$. 
$\nu_i$ and $D_i$ are the surface density of the lattice sites and the surface diffusivity, respectively. The surface chemical potentials read
\begin{eqnarray}
\mu_A&=&\frac{C_B^{(b)}}{\delta}\left(\frac{\partial \gamma}{\partial C_A}-\frac{\partial \gamma}{\partial C_B}\right)+\gamma \left(\frac{1}{F\sqrt{1+F_y^2}}-\frac{F_{yy}}{\left(1+F_y^2\right)^{3/2}}\right),\label{mu_A}\\
\mu_B&=&\frac{-\left(1-C_B^{(b)}\right)}{\delta}\left(\frac{\partial \gamma}{\partial C_A}-\frac{\partial \gamma}{\partial C_B}\right)+\gamma \left(\frac{1}{F\sqrt{1+F_y^2}}-\frac{F_{yy}}{\left(1+F_y^2\right)^{3/2}}\right), \label{mu_B}
\end{eqnarray}
where 
$\kappa=1/F\sqrt{1+F_y^2}-F_{yy}/\left(1+F_y^2\right)^{3/2}$ is
the surface mean curvature \cite{CFM,BBW}, and $\gamma$ the surface energy.  
Eqs. (\ref{mu_A}) and (\ref{mu_B}) result from the generalized thermodynamic forms derived 
in \cite{ZVD} after the surface energies are taken isotropic. 
This is justified by the fact that the surface energy anisotropy is much weaker for metals in comparison to 
semiconductors \cite{FLiu}. 

Since the binary alloys that we consider do not phase separate,
we use a simple interpolated form for $\gamma$,  
\begin{equation}
\gamma=\gamma_A C_A + \gamma_B C_B,\label{gamma}
\end{equation}
where $\gamma_A$ and $\gamma_B$ are the surface energies of a single-component wire that is 
made of atoms A or B \cite{RSK}.
Other expressions for $\gamma$ are possible, including adding terms comprising the ideal solution model 
to Eq. (\ref{gamma}) \cite{ZVD}, however this introduces
at least one additional parameter. Thus we avoided other expressions, aiming for the most basic mathematical model with 
as small as possible parameter space.  Eq. (\ref{gamma}) is necessary and sufficient to demonstrate how
different surface energies $\gamma_A$ and $\gamma_B$ manifest on PR instability through surface segregation. 

Last, the model is closed by stating $C_A+C_B=1$ (since the alloys are substitutional) and providing the expression for the equilibrium surface concentration:
\begin{equation}
C_B^{(eq)}=\frac{C_B^{(b)}}{\left(1-C_B^{(b)}\right)\exp{\left(\frac{\Delta E_{seg}}{k T}\right)}+C_B^{(b)}},
\; 
\Delta E_{seg}=\sigma_B\gamma_B-\sigma_A\gamma_A, \label{segreg_eq}
\end{equation}
where $\Delta E_{seg}$ is the enthalpy of surface segregation, and $\sigma_A$, $\sigma_B$ are the atomic areas. 
Eq. (\ref{segreg_eq}) is the classical Langmuir-McLean expression, which predicts surface segregation,  
$C_B^{(eq)}>C_B^{(b)}$, at $\Delta E_{seg}<0$. Thus the alloy component with a lower surface energy, $\gamma_B < \gamma_A$, 
segregates to the surface.   
$\Delta E_{seg}$ is related to interatomic binding energy of the single components of the alloy \cite{VSZ,VSZ1}.
The assumptions underlying our model formulation are in 
agreement with the assumptions of the  Langmuir-McLean model, i.e. an ideal 
and dilute bulk solution and a one atom thick surface layer \cite{WyK,Seah,RSK,LZH}. Dimensionless form of Eq. (\ref{segreg_eq}) is plotted 
in Fig. \ref{Fig1}. 

Let a deviation of the surface composition from its equilibrium value be described by a relative error-like condition
\begin{equation}
\frac{C_B-C_B^{(eq)}}{C_B^{(eq)}}=\beta,\quad \mbox{or}\; C_B=(1+\beta)C_B^{(eq)}. \label{reldev}
\end{equation}
Notice that $\beta>-1$ due to $C_B$ and $C_B^{(eq)}$ being positive. (At $\beta=-1$, $C_B=0$ 
(a single-component nanowire) and only a well-known morphology evolution equation 
$F_t=-\sqrt{1+F_y^2}\;{\mathbf \nabla}_s\cdot \Omega_A{\bf J}_A, {\bf J}_A=-M_A\gamma_A{\mathbf \nabla}_s \kappa$
needs to be retained \cite{CFM,BBW,McCVoorhees,WMVD,SVT,ZVD}.) 
The interval $-1<\beta<0$ corresponds to $C_B$ at the surface is less than an equilibrium value, whereas
the interval $\beta>0$ corresponds to $C_B$ at the surface is larger than an equilibrium value. 
Now, Eqs. (\ref{segreg_eq}) and (\ref{reldev}) provide:
\begin{eqnarray}
C_B^{(b)}&=&\frac{C_B}{\left(1+\beta-C_B\right)\exp{\left(\frac{-\Delta E_{seg}}{k T}\right)}+C_B}, \label{CB_bulk}\\
C_B^{(eq)}-C_B&=&\frac{-\beta}{1+\beta}C_B. \label{ka_term}
\end{eqnarray}
Next, $C_B^{(b)}$, $C_B^{(eq)}-C_B$, and $C_A$  are eliminated from the system 
(\ref{reqq})-(\ref{gamma}) using Eqs. (\ref{CB_bulk}), (\ref{ka_term}) and the condition of a substitutional alloy.
The coupled PDE system is now for $C_B$ and $F$. Further, 
using the time and length scales $t = \tau \bar t\equiv(\delta^4 k T/D_A\gamma_A\Omega_A^2\nu_A) \bar t$, $y=\delta \bar y$, $F=\delta \bar F$ we arrive to the final dimensionless PDE 
system


%
\begin{eqnarray}
\bar F_{\bar t}&=&\frac{1}{\sqrt{1+\bar F_{\bar y}^2}}\;\frac{\partial}{\partial \bar y}\left[\frac{\left(1-C_B\right)\frac{\partial \bar \mu_A}{\partial \bar y}+
M \sigma C_B\frac{\partial \bar \mu_B}{\partial \bar y}}{1+\bar F_{\bar y}^2}\right], \label{reqqq}\\
\frac{\partial C_B}{\partial \bar t}&=&\frac{-C_B}{\left(1+\bar F_{\bar y}^2\right)f\left(C_B\right)}\frac{\partial}{\partial \bar y}\left[\frac{\left(1-C_B\right)\frac{\partial \bar \mu_A}{\partial \bar y}+
M \sigma C_B\frac{\partial \bar \mu_B}{\partial \bar y}}{1+\bar F_{\bar y}^2}\right]
+ \frac{M \sigma}{1+\bar F_{\bar y}^2}\frac{\partial}{\partial \bar y} \left[\frac{C_B \frac{\partial \bar \mu_B}{\partial \bar y}}{1+\bar F_{\bar y}^2}\right]-\frac{\beta}{1+\beta}\epsilon\ C_B,\label{C_B_eqqq}
\end{eqnarray}
\begin{eqnarray}
\bar \mu_A&=&\frac{(1-\Gamma)C_B}{f\left(C_B\right)}+
\left(1-(1-\Gamma)C_B\right) \left(\frac{1}{\bar F\sqrt{1+\bar F_{\bar y}^2}}-\frac{\bar F_{\bar y \bar y}}{\left(1+\bar F_{\bar y}^2\right)^{3/2}}\right),\label{mu_A_ndim}\\
\bar \mu_B&=&-(1-\Gamma)\left(1-\frac{C_B}{f\left(C_B\right)}\right)+
\left(1-(1-\Gamma)C_B\right) \left(\frac{1}{\bar F\sqrt{1+\bar F_{\bar y}^2}}-\frac{\bar F_{\bar y \bar y}}{\left(1+\bar F_{\bar y}^2\right)^{3/2}}\right), \label{mu_B_ndim}
\end{eqnarray}
%


%
\begin{equation}
f\left(C_B\right)=\left(1+\beta-C_B\right)\exp{\left\{\alpha(1-\sigma \Gamma)\right\}}+C_B, \label{denominatorofCBb}
\end{equation}
where the parameters are: $M=M_B/M_A$ (the ratio of the surface mobilities), $\sigma=\sigma_B/\sigma_A$ (the ratio of the atomic areas),
$\Gamma=\gamma_B/\gamma_A$ (the ratio of the surface energies), $\alpha=\gamma_A \sigma_A/k T$ (the ratio of A-atom surface energy 
to the thermal energy), $\beta$ (the measure of the deviation from 
the equilibrium), and $\epsilon=K\tau/\delta$ (surface-bulk exchange rate of B-atoms). 
The last term on the right-hand side of 
Eq. (\ref{C_B_eqqq}) provides exponential growth of $C_B$ at $-1<\beta<0$, or exponential decay of $C_B$ at $\beta>0$. 

At $C_B=0$ (a single component nanowire) Eqs. (\ref{reqqq})-(\ref{denominatorofCBb}) reduce to a single equation for the morphology 
evolution \cite{NM2,BBW,CFM,WMVD}, 
\begin{equation}
\bar F_{\bar t}=\frac{1}{\sqrt{1+\bar F_{\bar y}^2}}\;\frac{\partial}{\partial \bar y}\left[\frac{\frac{\partial \bar \mu_A}{\partial \bar y}}{1+\bar F_{\bar y}^2}\right],\;
\bar \mu_A=\bar \kappa=\frac{1}{\bar F\sqrt{1+\bar F_{\bar y}^2}}-\frac{\bar F_{\bar y \bar y}}{\left(1+\bar F_{\bar y}^2\right)^{3/2}}. \label{primreq}
\end{equation}
At $\Gamma=1$ (equal surface energies of A and B atoms), 
$\bar \mu_A=\bar \mu_B=\bar \kappa$. 
At $\Gamma=1$ and $M\sigma=1$ (equal surface energies, equal surface mobilities, equal sizes of A and B atoms), the morphology evolution equation
reduces to Eq. (\ref{primreq}) and decouples from the concentration evolution equation; thus a morphology changes are
impacting a surface concentration, but not vice versa. Last, at $\sigma\Gamma=1$ (equal surface energies and equal sizes of A and B atoms), $f\left(C_B\right)=1+\beta$. 
At $\beta>-1$, $\beta\neq 0$ this
has the effect of changing the magnitudes of some terms in Eqs. (\ref{reqqq}) and (\ref{C_B_eqqq}) that contain 
$\partial C_B/\partial x$ and $\partial^2 C_B/\partial x^2$, while preserving the forms of these terms; the signature of a surface segregation is eliminated. 
At $\sigma\Gamma=1$ and $\beta=0$ a disequilibrium of a surface concentration is eliminated in addition to elimination 
of a surface segregation.
 
The typical interval of values for 
all dimensionless parameters, except $\epsilon$, is from 0.1 to 10.  The interesting value is $\epsilon$. 
Adopting the surface layer thickness 1 ML equal to the diameter of Ni atom 
($\delta = 0.25$ nm) and using other physical parameters (see Table \ref{T1}) result in $\tau\sim 10^{-9}$ s. 
Then using $K=2\times 10^{-11}$ cm$/$s \cite{VSZ} gives
$\epsilon \sim 10^{-12}$, and since $|\beta/(1+\beta)|$ is $O(1)$ or smaller, the exponential growth or decay of $C_B$ is extremely slow. 
Thus for LSA the term $\frac{-\beta}{1+\beta}\epsilon\ C_B$ is omitted from Eq. (\ref{C_B_eqqq}). This is also justified by the direct 
comparison of $|\frac{\beta}{1+\beta}|\epsilon\sim 10^{-12}$ to the perturbation growth rate $\omega \sim 10^{-5}-10^{-3}$ from LSA in 
Sec. \ref{LSA_res} for AuNi, PtNi, and CoNi wires.
The remaining dimensionless parameters in the model equations are $M$, $\sigma$, $\Gamma$, $\alpha$, and $\beta$, where $\beta$ 
is the free parameter. Values of $M$, $\sigma$, $\Gamma$, and $\alpha$ for three binary alloys 
are in Table \ref{T1}. 
In what follows, we also drop the overbars.

\section{LSA, Setup}
\label{LSA_set}

\begin{figure}[H]
\vspace{-0.2cm}
\centering
\includegraphics[width=3in]{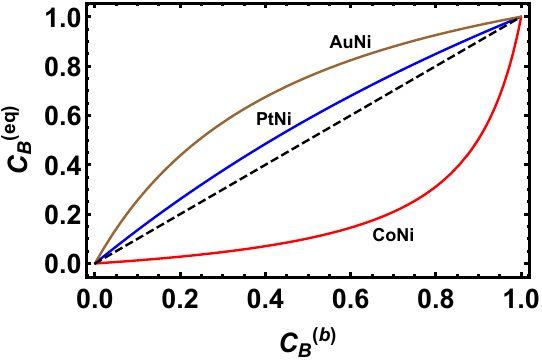}
\caption{Brown, blue, red curves: plots of Eq. (\ref{CB0base}) at $\beta=0$. Dashed line is the plot of equation $C_B^{(eq)}=C_B^{(b)}$.
}
\label{Fig1}
\end{figure}

Guided by the model formulation in Section \ref{Modell}, LSA will be performed about the physical homogeneous base state 
solution that corresponds to a wire of a constant radius 
$F(t,y)=R_0=r_0/\delta$ (as grown) that has the surface concentration 
\begin{equation}
C_B(t,y)=(1+\beta)C_B^{(eq)}=\frac{(1+\beta)C_B^{(b)}}{\left(1-C_B^{(b)}\right)\exp{\left\{-\alpha(1-\sigma \Gamma)\right\}}+C_B^{(b)}},\label{CB0base}
\end{equation}
ref. Eqs. (\ref{segreg_eq}) and (\ref{reldev}). 
Sub-surface concentration $C_B^{(b)}$ mimics the bulk composition of a wire, and it is yet another input parameter to the model. 
Fig. \ref{Fig1} shows that over the entire 
unit interval of the bulk concentration, the equilibrium surface concentration exceeds the bulk concentration for AuNi and PtNi, 
but it is significantly smaller than the bulk one for CoNi. Notice that at this point the set of all dimensionless input parameters is 
${\mathbf p_a}=\left\{M, \sigma, \Gamma, \alpha, \beta, R_0, C_B^{(b)}\right\}$. We also introduce a reduced parameter set
${\mathbf p_r}=\left\{M, \sigma, \Gamma, \alpha, C_B^{(b)}\right\}$.

Introduce perturbations of the base state, $F(t,y)=R_0+f(t,y)$, $C_B(t,y)=(1+\beta)C_B^{(eq)}+c(t,y)$, $|f(t,y)|, |c(t,y)|\ll 1$, substitute in
Eqs. (\ref{reqqq})-(\ref{denominatorofCBb}) (with the last term in Eq. (\ref{C_B_eqqq} omitted) and linearize. This results in the equations for the perturbations that have the form
$f_t=g\left(c_{yy},f_{yy},f_{yyyy}\right)$, $c_t=h\left(c_{yy},f_{yy},f_{yyyy}\right)$,
where $g$ and $h$ are a linear functions of their arguments. Next, substituting the normal modes $f(t,y)=u\exp{(\omega t + iky)}$,
$c(t,y)=v\exp{(\omega t + iky)}$ yields an algebraic linear homogeneous system for the amplitudes $u$ and $v$. The condition for existence 
of a non-zero solution is that the determinant of the system's matrix is equal to zero, which results in the typical dispersion relation
for the long-wavelength instability
that has the form $\omega(k)=\theta_2\left({\mathbf p_a}\right) k^2 - \theta_4\left({\mathbf p_r}\right) k^4$, where 
$\theta_4\left({\mathbf p_r}\right)=\left(1+(1+\beta)(\Gamma-1)C_B^{(eq)}\right)\left(1+(1+\beta)(M \sigma-1)C_B^{(eq)}\right)$ and $\theta_2\left({\mathbf p_a}\right)$ depends in a complicated way 
on all parameters. 
($\theta_2\left({\mathbf p_a}\right)$ has the form 
$P_5\left((1+\beta)C_B^{(eq)}\right)/P_3\left((1+\beta)C_B^{(eq)}\right)$, where 
$P_3$, $P_5$ 
are the polynomials of degree three and five, respectively, with a cumbersome coefficients). The cutoff and the most dangerous perturbation wavelengths are
\begin{equation}
\lambda_c=\frac{2 \pi}{k_c}=\frac{2 \pi}{\sqrt{\frac{\theta_2\left({\mathbf p_a}\right)}{\theta_4\left({\mathbf p_r}\right)}}}= 2\pi \sqrt{\frac{\theta_4\left({\mathbf p_r}\right)}{\theta_2\left({\mathbf p_a}\right)}},\; \lambda_{max}=\sqrt{2}\ \lambda_c \label{lambdas}.
\end{equation}
Here $k_c$ is the cutoff wavenumber of the long-wavelength instability. $\omega(k)$  
reduces to $\omega(k)=k^2\left(1/R_0^2-k^2\right)$ (a single-component case \cite{NM2,McCVoorhees}) at  
$\Gamma=1$, and either $\beta=-1$, or $M\sigma=1$. 

An axial perturbation of a wire surface (or an axial perturbation of an equilibrium surface concentration) with a wavelength larger 
than $\lambda_c$ destabilizes the base state, leading ultimately to a wire breakup into an alloy particles, 
in fact a short cylindrical segments, with a spacing 
$\lambda_{max}$ between the segment's centers 
and a surface concentration that may be distinctly different 
from the equilibrium one. From this picture it is clear that a particle diameter (a length of a segment) also can be 
approximated as $\lambda_{max}$. Since a choice of a binary alloy fixes the parameters $M$, $\sigma$, $\Gamma$ and $\alpha$, 
the set of free dimensionless parameters is $\left\{\beta, R_0, C_B^{(b)}\right\}$.
Below we investigate a linear instability of the base state as a function of these parameters.
In all figures below we cite a physical radius $r_0$ instead of $R_0$ for transparency. 

\begin{figure}[H]
\vspace{-0.2cm}
\centering
\includegraphics[width=5.0in]{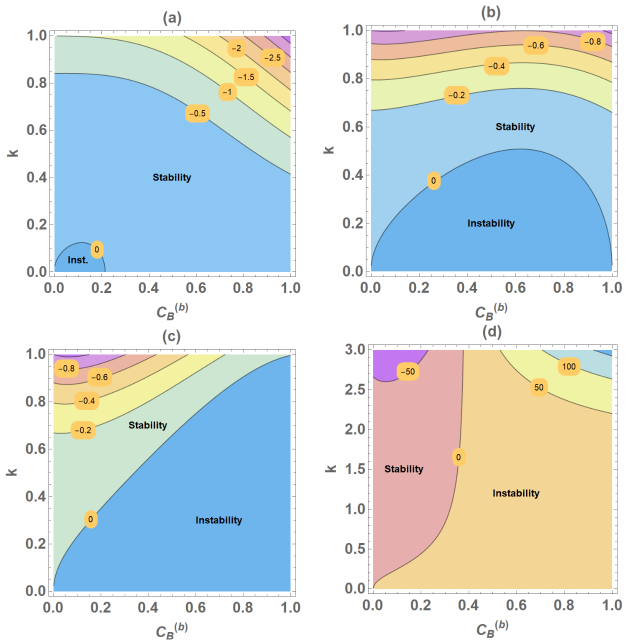}
\caption{AuNi: Contour plot of $\omega$ at $r_0=10$ nm, and (a) $\beta=-0.7$, (b) $\beta=0$, (c) $\beta=1$, (d) $\beta=3$.
}
\label{Fig2}
\end{figure}

Before presenting LSA results in a uniform fashion for three binary alloys in the next section, Fig. \ref{Fig2} shows
the plots of the perturbation growth rate  $\omega$ only for AuNi, as these plots are qualitatively similar 
for the other two alloys.
There is a longwave instability at all bulk concentration values for moderately large positive $\beta$ and for 
an interval of (relatively) small bulk concentrations for negative $\beta$ (Fig. \ref{Fig2}(a-c)). 
The instability domain grows as $\beta$ increases from $-1$, i.e. as $C_B$ increases from zero and surpasses 
its equilibrium value (ref. Eq. (\ref{reldev})). The cap-like shape of the instability domain for fairly small $|\beta|$ points to a near 
zero and (relatively) large $C_B^{(b)}$ cases deliver a large instability cutoff wavelength $\lambda_c$ and therefore, 
a large $\lambda_{max}$. 
At $\beta>0$ and not exceeding a threshold these quantities are large at small $C_B^{(b)}$ and they decrease as 
$C_B^{(b)}$ increases (Fig. \ref{Fig2}(c)). When $\beta$ crosses a positive threshold value $\beta_{th}$, i.e. a deviation from equilibrium 
is too large, only the cases with 
a relatively small $C_B^{(b)}$ remain physical, i.e. there still is a short-wavelength cutoff of unstable modes 
(Fig. \ref{Fig2}(d)).
A typical $\omega_{max}=\omega\left(k_{max}\right)$ is several hundred to several thousands times 
larger than $\omega_{max}=k_{max}^2\left(1/R_0^2-k_{max}^2\right)$ for a single-component nanowire. 
This is true also for PtNi and CoNi nanowires.

\section{LSA, Results}
\label{LSA_res}

\begin{figure}[H]
\vspace{-0.2cm}
\centering
\includegraphics[width=5.0in]{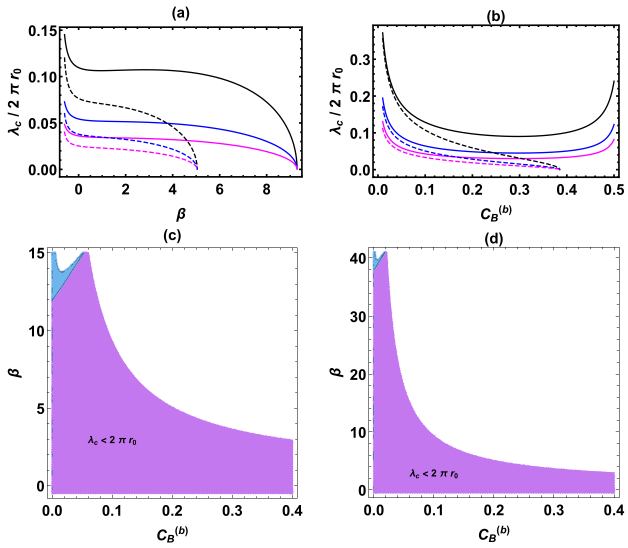}
\caption{AuNi: (a) Solid lines: $C_B^{(b)}=0.1$, dashed lines: $C_B^{(b)}=0.2$; Black, blue, magenta line colors: $r_0=10, 20, 30$ nm, respectively. 
Plot interval $-0.6\le \beta \le 9.3$.
(b) Solid lines: $\beta=-0.5$, dashed lines: $\beta=3$; Black, blue, magenta line colors: $r_0=10, 20, 30$ nm, respectively.
(c,d): Domains in the $C_B^{(b)}\beta$ plane, where $\lambda_c<2\pi r_0$ (magenta)
or  $\lambda_c>2\pi r_0$ (blue).
(c) $r_0 =10$ nm, $\lambda_c=2\pi r_0$ boundary equation: $11.87+59.68 C_B^{(b)}$, $0<C_B^{(b)}\le 0.06$.  
(d) $r_0 =30$ nm, $\lambda_c=2\pi r_0$ boundary equation: $37.83+132.36 C_B^{(b)}$, $0<C_B^{(b)}\le 0.025$. 
}
\label{Fig3}
\end{figure}

For AuNi Figures \ref{Fig3}(a,b) show that the ratio of the cutoff wavelength to $2\pi r_0$ is less than one, unless 
$\beta\rightarrow \beta_c+,\; \beta_c<0$, or $C_B^{(b)}\rightarrow 0+$ at $\beta<>0$, 
or $C_B^{(b)}\rightarrow \tilde C_B^{(b)}-$ at $\beta<0$. For instance, in Fig. \ref{Fig3}(a,b) $\tilde C_B^{(b)}=0.5$ at $\beta=-0.5$.
The curves in Fig. \ref{Fig3}(a) terminate at a positive threshold $\beta_{th}$, since beyond
a threshold (large deviations from equilibrium) the model loses validity. $\beta_{th}$ decreases as $C_B^{(b)}$ increases. 
Same is true for dashed curves ($\beta=3$)
in Fig. \ref{Fig3}(b). Larger $C_B^{(b)}$ or larger wire radius (Fig. \ref{Fig3}(a)) or larger $\beta$ (Fig. \ref{Fig3}(b)) 
result in smaller ratio.
Figures \ref{Fig3}(c,d) show the domains in the $C_B^{(b)}\beta$ plane, where $\lambda_c<2\pi r_0$ 
or  $\lambda_c>2\pi r_0$, for two values of a wire radius. At fixed $C_B^{(b)}>C_{B,c}^{(b)}$ the former outcome occurs for 
$\beta<\beta_{th}$. $\beta_{th}$ is in inverse power relationship to $C_B^{(b)}$, but the exponent is difficult to 
determine. 
At $\beta>\beta_{th}$ the model is not valid (white areas). At $C_B^{(b)}<C_{B,c}^{(b)}$, $\lambda_c<2\pi r_0$  
at $\beta$ below the critical line that separates the magenta and the blue region, where  $\lambda_c>2\pi r_0$.
$C_{B,c}^{(b)}$ decreases 
as the radius increases: $C_{B,c}^{(b)}=0.06$ at $r_0=10$ nm, and $C_{B,c}^{(b)}=0.025$ at $r_0=30$ nm. 
Fits to the critical line were found, see Fig. \ref{Fig3} caption.

\begin{figure}[H]
\vspace{-0.2cm}
\centering
\includegraphics[width=5.0in]{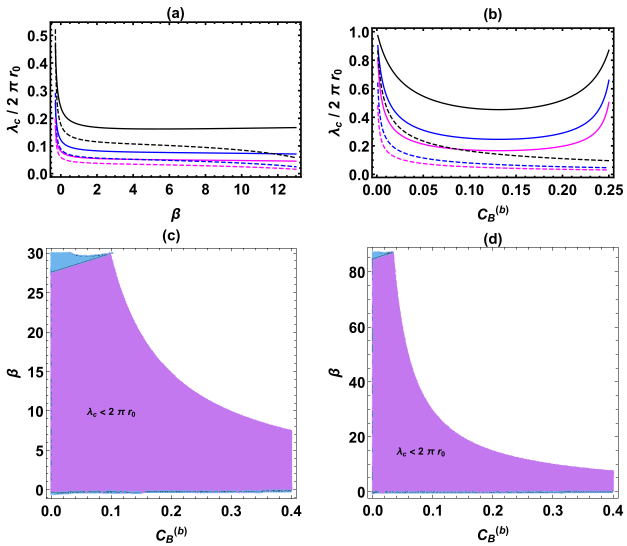}
\caption{PtNi: (a) Solid lines: $C_B^{(b)}=0.1$, dashed lines: $C_B^{(b)}=0.2$; Black, blue, magenta line colors: $r_0=10, 20, 30$ nm, respectively. 
Plot interval $-0.3\le \beta \le 13$.
(b) Solid lines: $\beta=-0.3$, dashed lines: $\beta=3$; Black, blue, magenta line colors: $r_0=10, 20, 30$ nm, respectively. 
(c,d): Domains in the $C_B^{(b)}\beta$ plane, where $\lambda_c<2\pi r_0$ (magenta)
or  $\lambda_c>2\pi r_0$ (blue).
(c) $r_0 =10$ nm, $\lambda_c=2\pi r_0$ boundary equation: $27.57+23.84 C_B^{(b)}$, $0<C_B^{(b)}\le 0.11$.  
(d) $r_0 =30$ nm, $\lambda_c=2\pi r_0$ boundary equation: $84.72+70.96 C_B^{(b)}$, $0<C_B^{(b)}\le 0.05$. 
}
\label{Fig4}
\end{figure}

Results for PtNi (Fig. \ref{Fig4}) are expectedly qualitatively similar to the ones for AuNi due to the similarity of the equilibrium curves 
in Fig. \ref{Fig1}. 
The ratio $\lambda_c/2\pi r_0$ in Figures \ref{Fig4}(a,b) shows the same trends  as in 
Figures \ref{Fig3}(a,b). However, the ratio values are four to six times larger. Another significant difference 
in Figures \ref{Fig4}(a,b) is that $\beta_{th}$ values 
are larger and do not fit into the optimal plot window.
Correspondingly in Figures \ref{Fig4}(c,d) the domain $\lambda_c< 2\pi r_0$ is larger, i.e. the model validity boundary is 
at larger $\beta$ values. The domain $\lambda_c> 2\pi r_0$ also is at larger $\beta$ and it extends over 
wider interval of bulk concentration values: 
$C_{B,c}^{(b)}=0.11$ in Fig. \ref{Fig4}(c) and $C_{B,c}^{(b)}=0.05$ in Fig. \ref{Fig4}(d). 

\begin{figure}[H]
\vspace{-0.2cm}
\centering
\includegraphics[width=5.0in]{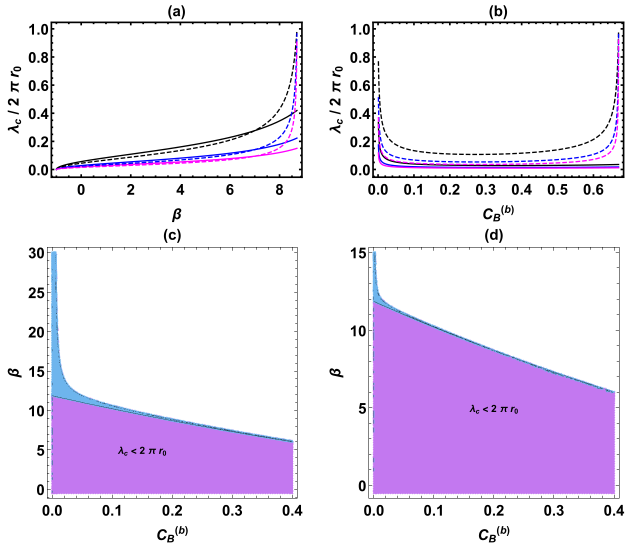}
\caption{CoNi: (a) Solid lines: $C_B^{(b)}=0.1$, dashed lines: $C_B^{(b)}=0.2$; Black, blue, magenta line colors: $r_0=10, 20, 30$ nm, respectively. 
Plot interval $-0.99\le \beta \le 8.7$.
(b) Solid lines: $\beta=-0.5$, dashed lines: $\beta=3$; Black, blue, magenta line colors: $r_0=10, 20, 30$ nm, respectively. 
(c,d): Domains in the $C_B^{(b)}\beta$ plane, where $\lambda_c<2\pi r_0$ (magenta)
or  $\lambda_c>2\pi r_0$ (blue).
(c) $r_0 =10$ nm, $\lambda_c=2\pi r_0$ boundary equation: $11.83-15.88 C_B^{(b)}$.  
(d) $r_0 =30$ nm, $\lambda_c=2\pi r_0$ boundary equation: $11.76-15.22 C_B^{(b)}$. 
}
\label{Fig5}
\end{figure}

Fig. \ref{Fig5} shows that the results for CoNi are quite different from the ones for AuNi and PtNi. 
In Fig. \ref{Fig5}(a) $\lambda_c/2\pi r_0\approx 0$ at $\beta$ slightly larger than -1 and the ratio increases as $\beta$ 
approaches $\beta_{th}$. The latter value again decreases as $C_B^{(b)}$ increases. 
The ratio $\lambda_c/2\pi r_0$ is not sensitive to a radius changes. In Fig. \ref{Fig5}(b) the dashed curves 
(large positive $\beta$) are similar to the solid curves 
(negative $\beta$) in Figures \ref{Fig3}(b), \ref{Fig4}(b) and vice versa.
The shape and size of the $\lambda_c<2\pi r_0$ region in
Figures \ref{Fig5}(c,d) is also different from Figures \ref{Fig3} and \ref{Fig4}. The domain is trapezoidal and its upper 
boundary everywhere separates from the  domain $\lambda_c>2\pi r_0$, that is, for every $C_B^{(b)}$ that is not too close 
to 1 there is a $\beta$ value such that $\lambda_c>2\pi r_0$. $\lambda_c=2\pi r_0$ line has a negative slope, and overall
values of $\beta$ that deliver $\lambda_c<2\pi r_0$ are 1.5 to 2 times smaller than for AuNi or PtNi.

\begin{table}[!ht]
\centering
{\scriptsize 
\begin{tabular}
{|c|c|c|c|c|c|}

\hline
			\rule[-2mm]{0mm}{6mm} \textbf{AuNi} & \textbf{AuNi} & \textbf{PtNi} & \textbf{PtNi} & \textbf{CoNi} & \textbf{CoNi}	 \\
			\hline	 
			\rule[-2mm]{0mm}{6mm} \textbf{Parameter} & \textbf{Parameter} & \textbf{Parameter} & \textbf{Parameter} & \textbf{Parameter} & \textbf{Parameter}	 \\
			\hline
            \hline

			    \rule[-2mm]{0mm}{6mm} $\gamma_A=1850$ erg/cm$^2$ & $\Gamma=0.62$ & $\gamma_A=1850$ erg/cm$^2$ & $\Gamma=0.76$ & $\gamma_A=1850$ erg/cm$^2$ & $\Gamma=1.29$ \\ 
				\hline
				\rule[-2mm]{0mm}{6mm} $\gamma_B=1150$ erg/cm$^2$ & & $\gamma_B=1400$ erg/cm$^2$ & & $\gamma_B=2390$ erg/cm$^2$ & \\ 
				\hline
				\rule[-2mm]{0mm}{6mm} $\Omega_A=7.98\times 10^{-24}$ cm$^3$ & $M=1.26$ & $\Omega_A=7.98\times 10^{-24}$ cm$^3$ & $M=0.92$ & $\Omega_A=7.98\times 10^{-24}$ cm$^3$ & $M=0.67$\\ 
				\hline
				\rule[-2mm]{0mm}{6mm} $\Omega_B=1.25\times 10^{-23}$ cm$^3$ & & $\Omega_B=1.12\times 10^{-23}$ cm$^3$ & & $\Omega_B=8.18\times 10^{-24}$ cm$^3$ & \\
				\hline
				\rule[-2mm]{0mm}{6mm} $\sigma_A=4.83\times 10^{-16}$ cm$^2$ & $\sigma=1.35$ & $\sigma_A=4.83\times 10^{-16}$ cm$^2$ & $\sigma=1.26$ & $\sigma_A=4.83\times 10^{-16}$ cm$^2$ & $\sigma=1.02$\\ 
				\hline
				\rule[-2mm]{0mm}{6mm} $\sigma_B=6.51\times 10^{-16}$ cm$^2$ & & $\sigma_B=6.07\times 10^{-16}$ cm$^2$ & & $\sigma_B=4.91\times 10^{-16}$ cm$^2$ & \\ 
				\hline
				\rule[-2mm]{0mm}{6mm} $\nu_A=1.3\times 10^{15}$ cm$^{-2}$ & $\alpha=7.02$ & $\nu_A=1.3\times 10^{15}$ cm$^{-2}$ & $\alpha=7.02$ & $\nu_A=1.3\times 10^{15}$ cm$^{-2}$ & $\alpha=7.02$ \\ 
				\hline
				\rule[-2mm]{0mm}{6mm} $\nu_B=9.9\times 10^{14}$ cm$^{-2}$ & & $\nu_B=1.07\times 10^{15}$ cm$^{-2}$ & & $\nu_B=1.3\times 10^{15}$ cm$^{-2}$ & \\ 
				\hline
				\rule[-2mm]{0mm}{6mm} $D_A=5\times 10^{-7}$ cm$^2$/s & & $D_A=5\times 10^{-7}$ cm$^2$/s & & $D_A=5\times 10^{-7}$ cm$^2$/s & \\ 
				\hline
				\rule[-2mm]{0mm}{6mm} $D_B=5.3\times 10^{-7}$ cm$^2$/s & &  $D_B=4\times 10^{-7}$ cm$^2$/s & & $D_B=3.3\times 10^{-7}$ cm$^2$/s & \\ 
				\hline
				\rule[-2mm]{0mm}{6mm}  $\delta=0.248$ nm & & $\delta=0.248$ nm & & $\delta=0.248$ nm & \\
				\hline
				\rule[-2mm]{0mm}{6mm}   $K=2\times 10^{-11}$ cm/s & & $K=2\times 10^{-11}$ cm/s & & $K=2\times 10^{-11}$ cm/s &   \\
				\hline
\end{tabular}}
\caption[\quad Parameters]{Physical and dimensionless parameters.
}
\label{T1}
\end{table}

\section{Conclusions}
\label{Summary}

\begin{figure}[H]
\vspace{-0.2cm}
\centering
\includegraphics[width=5.0in]{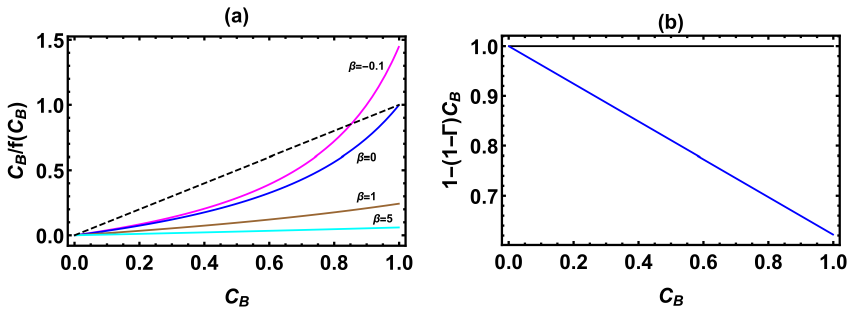}
\caption{AuNi. (a) Colored curves: plots of $C_B/f\left(C_B\right)$  at various $\beta$. 
Dashed line is the plot of $C_B$, i.e. the ratio is taken at $f\left(C_B\right)=1$. (b) Plot of $1 - (1 - \Gamma)C_B$ (blue line).
}
\label{Fig6}
\end{figure}

Smallness of the critical wavelength in the presented minimal model is attributed primarily to surface segregation. Account of the latter 
resulted in $C_B$ replaced by $C_B/f\left(C_B\right)$ in the first terms on the right-hand side of the dimensionless Eqs. (\ref{C_B_eqqq})-(\ref{mu_B_ndim}). 
For AuNi as the example, Fig. \ref{Fig6}(a) shows that 
division by $f\left(C_B\right)$, see Eq. (\ref{denominatorofCBb}) changes $C_B$ very strongly in the whole of the composition interval. 
The quotient deviates significantly from $C_B$ even at equilibrium, $\beta=0$ (the maximum deviation is 43\% at $C_B\approx 0.64$).
In addition, Fig. (\ref{Fig6})(b) shows that the factor $1 - (1 - \Gamma)C_B$ that multiplies the surface mean curvature in the surface chemical potentials 
(\ref{mu_A_ndim}) and (\ref{mu_B_ndim}), at $C_B>0$ and 
$\Gamma\neq 1$ differs from its value 1 at $C_B=0$ (a single-component wire) or at $C_B>0$ and $\Gamma=1$ (equal surface energies of the components A and B), thus enhancing the impact
of a surface segregation on the critical wavelength.

In future, the model will be expanded in order to understand which physical factors may 
increase the critical wavelength to cross the classical circumference threshold. Critical assessment of the literature will
be undertaken, and next the model will be enriched by  
including a compositional stress \cite{SVT,KH} and if necessary, 
one or more effects that may manifest strongly in a binary alloy nanowire that is annealed at high temperature, such as
spinodal decomposition or surface directed spinodal decomposition (at temperatures below 550$^\circ$), contact lines with a substrate, and external forces (electromigration). 

\appendix

\section{Remarks on Eq. (\ref{C_B_eqq})}
\label{BasicEqs}

Eq. (\ref{C_B_eqq}) follows from the PDE derived in Ref. \cite{ZVD} using mass conservation (see Eq. (1) and Appendix A in Ref. \cite{ZVD}).
Omitting the deposition term that is not relevant to evolution under the annealing conditions, that PDE reads:  
\begin{equation}
\delta\frac{\partial C_B}{\partial t}=-C_B^{(b)}V_n-\Omega_B{\mathbf \nabla}_s\cdot{\bf J}_B+
\Omega_B\left(1-C_B^{(b)}\right){\mathbf n}\cdot {\mathbf J}_B^{(b)} - \Omega_A C_B^{(b)}{\mathbf n}\cdot {\mathbf J}_A^{(b)}, \label{C_B_eq_large}
\end{equation}
where ${\mathbf n}$ is a unit normal to the surface, and ${\mathbf J}_A^{(b)}, {\mathbf J}_B^{(b)}$ are the fluxes from the 
bulk. Eq. (\ref{C_B_eq_large}) was also derived, with minor differences, by Spencer, Voorhees and Tersoff in the earlier fundamental paper \cite{SVT}. 

Taking $C_B^{(b)}\ll 1$ (a dilute alloy) reduces the last two terms in Eq. (\ref{C_B_eq_large}) to $\Omega_B{\mathbf n}\cdot {\mathbf J}_B^{(b)}$, and then 
adopting ${\mathbf n}\cdot {\mathbf J}_B^{(b)}=K\left(C_B^{(eq)}-C_B\right)$ \cite{VSZ}
yields Eq. (\ref{C_B_eqq}). When the surface and the bulk chemical potentials differ, the 
flux ${\mathbf J}_B^{(b)}$ is not zero and tends to restore the thermodynamic equilibrium between bulk and surface. The assumption of a dilute alloy is consistent with the surface segregation model, 
Eq. (\ref{segreg_eq}). Note that if, further, $C_B^{(b)}=0$ (the bulk is absent), then Eq. (\ref{C_B_eqq}) reduces to a 
diffusion of an adsorbate of a thickness $\delta$ on a stationary planar substrate, $\delta\frac{\partial C_B}{\partial t}=-\Omega_B{\mathbf \nabla}_s\cdot{\bf J}_B$. 

Equation that is nearly identical to Eq. (\ref{C_B_eqq}) is stated and employed, without derivation, by Vastola \textit{et al.} \cite{VSZ,VSZ1}. 
Their version of Eq. (\ref{C_B_eqq}) has $C_B^{(b)}$ replaced by $C_B$ in the first term on the right-hand side, and the 
second term $-\Omega_B{\mathbf \nabla}_s\cdot{\bf J}_B$ is absent. They are not stating the chemical potentials.
We repeated all calculations in the paper using Vastola \emph{et al.} version of Eq. (\ref{C_B_eqq}), still with the surface chemical potentials 
(\ref{mu_A}) and (\ref{mu_B}), and found that the results are 
qualitatively very similar to the presented ones. Using Vastola \emph{et al.} version typically results in the ratios $\lambda_c/2\pi r_0$ that are 
around two times larger.


\end{document}